\documentclass{article}

\usepackage{amsmath}  
\usepackage[english]{babel}
\usepackage[utf8]{inputenc}
\usepackage{amssymb}
\usepackage{graphicx}
\usepackage{authblk}
\usepackage{caption}
\usepackage{multirow}
\usepackage{amscd}
\usepackage{geometry}
\usepackage{a4wide}
\usepackage[usenames]{color}
\usepackage{etaremune} 
\usepackage{amsfonts}
\usepackage{ifthen}
\usepackage{color}
\usepackage{amscd}
\usepackage{latexsym}
\usepackage[hidelinks]{hyperref}

\newcommand{\bd}{\begin{definition}}
	\newcommand{\ed}{\end{definition}}
\newcommand{\bt}{\begin{theorem}}
	\newcommand{\et}{\end{theorem}}
\newcommand{\bi}{\begin{itemize}}
	\newcommand{\ei}{\end{itemize}}
\newcommand{\ben}{\begin{enumerate}}
	\newcommand{\een}{\end{enumerate}}
\newcommand{\beq}{\begin{equation}}
\newcommand{\eeq}{\end{equation}}

\newcommand{\R}{\mbox{$ \mathbb{R}  $}}

\newtheorem{definition}{Def.}[section]
\newtheorem{theorem}{Theorem}[section]

\def \proof{\noindent{\it Proof. \;}  \ignorespaces}
\def \qed{ \hfill $\Box$ \\}

\def \proof{\noindent{\it Proof}.  \ignorespaces}
\def \qed{ \hfill $\Box$ \\}

\begin{document}

\title{The quantum nature of color perception: Uncertainty relations for chromatic opposition}
\author[1]{Michel Berthier\thanks{michel.berthier@univ-lr.fr}}
\author[2]{Edoardo Provenzi\thanks{edoardo.provenzi@math.u-bordeaux.fr}}
\affil[1]{Laboratoire MIA, La Rochelle Université, Avenue Albert Einstein, BP 33060, 17031 La Rochelle, France}
\affil[2]{Université de Bordeaux, CNRS, Bordeaux INP, IMB, UMR 5251. F-33400, 351 Cours de la Libération}

\renewcommand\Authands{ and }
\date{}

\maketitle


\begin{abstract}
In this paper we provide an overview on the foundation and first results of a very recent quantum theory of color perception, together with novel results about uncertainty relations for chromatic opposition. The major inspiration for this model is the 1974 remarkable work by H.L. Resnikoff, who had the idea to give up the analysis of the space of perceived colors through metameric classes of spectra in favor of the study of its algebraic properties. This strategy permitted to reveal the importance of hyperbolic geometry in colorimetry. Starting from these premises, we show how Resnikoff's construction can be extended to a geometrically rich quantum framework, where the concepts of achromatic color, hue and saturation can be rigorously defined. Moreover, the analysis of pure and mixed quantum chromatic states leads to a deep understanding of chromatic opposition and its role in the encoding of visual signals. We complete our paper by proving the existence of uncertainty relations for the degree of chromatic opposition, thus providing a theoretical confirmation of the quantum nature of color perception.
\end{abstract}


\section{Introduction}

The central core of this paper is the concept of space of colors perceived by a trichromatic human being, or color space, for short. The scientific literature about this topic is abundant and here we limit ourselves to quote the classical reference \cite{Wyszecky:82}, the more mathematically-oriented books \cite{Dubois:09} and \cite{Koenderink:2010} and the image processing and computer vision oriented references \cite{Ebner:07}  and \cite{Gevers:12}, among hundreds of books and papers written on this subject.

What makes scientists so interested in color spaces is that, instead of being simple collections of elements  representing either physical color stimuli or the sensation they provoke in humans, they are \textit{structured spaces}, with intrinsic algebraic and geometrical properties and a metric able to quantify the distance (physical or perceptual) between their points.

The work of H.L. Resnikoff, that we will briefly recall in section \ref{sec:Y&R}, is related to the aforementioned geometrical and metric structure of the perceived color space. It is a foundational work, without direct algorithmic applications and it requires a non trivial acquaintance with theoretical physics and pure mathematics. This is probably the reason why his contribution, that we consider so important, has remained practically unnoticed until now. 

In this paper we start providing an outlook on the color perception theory inspired by Resnikoff's insights that we have developed through the papers  \cite{BerthierProvenzi:19,Provenzi:20,Berthier:2020,Prencipe:20,Berthier:21JMIV,Berthier:21JMP}. These are quite technical and dense works, here we prefer to privilege clarity of exposition and, for this reason, we will omit the proofs of the results that we will claim, the interested reader may consult them in the contributions just quoted.

After introducing, in section  \ref{sec:Jalgebras}, some fundamental mathematical results related to Jordan algebras and their use in quantum theories, we show, in section \ref{sec:quantum}, that Resnikoff's work leads to a quantum theory of color perception. As a novel contribution, we will provide an extensive motivation to explain why we believe that a quantum theory is better suited than a classical one to model color perception and we will also provide a strong theoretical support for our claim, i.e. the existence of uncertainty relations for color opponency.

In section \ref{sec:quantstates} we show how to identify pure and mixed quantum chromatic states, this result is used in section \ref{sec:vonNeumannsat} to study hue and saturation of a color and in section \ref{sec:Hering} to show that the peculiar achromatic plus chromatic opponent encoding of light signals performed by the human visual system can be intrinsically described by the quantum framework, without resorting to an `a-posteriori' statistical analysis. In section \ref{sec:uncertainty} we derive the uncertainty relations for chromatic opponency by adapting a technique first proposed by Schrödinger that extends the one used by Heisenberg, Weyl and Robertson to single out uncertainty bounds. Section \ref{sec:geometryofQstates} offers a brief panorama on the rich geometry of quantum chromatic states. We conclude our paper in section \ref{sec:discussion} by discussing some ideas about future developments of the theory.

\section{The dawn of hyperbolicity in color perception: Yilmaz's and Resnikoff's works}\label{sec:Y&R}
Here we present, in two separate subsections, the ideas and results of H. Yilmaz and H.L. Resnikoff about the role of hyperbolic structures in the study of color perception. In spite of the fact that Resnikoff's contribution is, by far, more important than Yilmaz's for our purposes, we want to respect the chronological development of the two works and we start with Yilmaz's idea about the link between color perception and special relativity.

\subsection{Yilmaz's relativity of color perception}\label{sec:Yilmaz}
Yilmaz was a theoretical physicist specialized in relativity who, around 1960, started to apply his knowledge to the field of human perception, here we shortly recap his 1962 paper \cite{Yilmaz:62} about the relativity of color perception. A detailed version of the content of this section can be found in \cite{Prencipe:20}. 

Yilmaz considered a visual scene where a trichromatic observer adapted to a broadband illuminant can identify the colors of a patch by performing comparison with a set of Munsell chips. He searched for the transformation that permits to relate the color description when the observer is adapted to two different broadband illuminants. As a first approximation, he searched for a linear map, i.e. a matrix transformation between color coordinated, and he derived the explicit form of the entries of this matrix by using the result of three perceptual experiments. He obtained a three-dimensional Lorentzian matrix, with Lorentz factor given by $\Gamma=1/\sqrt{1-(\sigma/\Sigma)^2}$, in which the perceived saturation $\sigma$ and the maximal perceivable saturation $\Sigma$ are the analogous of the speed of a moving particle and the speed of light, respectively, in special relativity. 

%
%

Lorentz transformations are precisely the linear maps that preserve the Lorentzian scalar product, which is the hyperbolic counterpart of the Euclidean scalar product, see e.g.  \cite{Ratcliffe:2006}. Thus, up to our knowledge, Yilmaz underlined for the first time that hyperbolic structures may play a significant role in color perception.

Yilmaz ideas, surely brilliant and much ahead of his time, were developed only heuristically: his mathematical analysis is not fully convincing in multiple parts and the experimental results that allow him to build the Lorentz transformation are just claimed, without providing any experimental data or apparatus description, see \cite{Prencipe:20} and \cite{Berthier:21JMP} for further details. In spite of that, Yilmaz's hint on the importance of hyperbolic geometry for the study of color perception inspired at least one key scientist, the polymath H.L. Resnikoff, who acknowledged Yilmaz in his fundamental 1974 paper \cite{Resnikoff:74} that we recall in the next subsection.

\subsection{Resnikoff's homogeneous color space}\label{sec:Resnikoff}

To greet Resnikoff's foresight, we consider that no sentence is more appropriate than that of Altiero Spinelli, one of the founding fathers of the European Union, who stated the following: `\textit{the quality of an idea is revealed by its ability to rise again from its failures}'. Resnikoff's work was written in the language of mathematical physics and fused differential geometry, harmonic analysis of Lie groups and the theory of Jordan algebras to study geometry and metrics of the space of perceived colors. This was far too abstract and advanced for the typical mathematical knowledge of the average color scientist of that time, with the consequence that Resnikoff's paper failed to interest the scientific community until today.

Resnikoff's cleverest idea was to abandon the classical description of perceived colors in terms of metameric classes of light spectra, see e.g. \cite{Wyszecky:82}, and to concentrate on an alternative description based, essentially, on \textit{the algebraic properties satisfied by perceived colors}. As we will see, this turned out to be the key to unveil a completely new way of representing colors.

The starting point of Resnikoff's paper is the beautiful 1920 Schrödinger's set of axioms of perceived colors: the great theoretical physicist, before dedicating himself to quantum mechanics, studied optics and color perception and came to the conclusion that the empirical discoveries of the founding fathers of color theory, none less than Newton \cite{Newton:52,Newton:93}, Helmholtz \cite{Helmholtz:05}, Grassmann \cite{Grassmann:1853} and Maxwell \cite{Maxwell:1857}, could be resumed in a set of axioms which, put together, say that \textit{the space of perceived colors}, denoted with $\cal C$ from now on, \textit{is a regular convex cone of dimension 3} (for trichromatic observers, the only ones that we consider in this paper).

The mathematical formalization of this concept is the following: let $\cal C$ be a subset of a finite dimensional inner product vector space $(V,\langle \, ,\, \rangle)$, then:
\begin{itemize}
	\item $\cal C$ is a \textit{cone} if, for all $c\in \cal C$ and all $\lambda \in \R^+$, $\lambda c\in \cal C$, i.e. $\cal C$ is stable w.r.t. multiplications by a positive constant.  This is the mathematical translation of the fact that, up to the glare limit, if we can perceive a color, then we can also perceive a brighter version of it;
	\item $\cal C$ is \textit{convex} if, for every couple $c_1,c_2 \in \cal C$ and every $\alpha \in [0,1]$, $\alpha c_1+(1-\alpha)c_2 \in \cal C$, i.e. the line segment connecting two perceived colors is entirely composed by perceived colors;
	\item $\cal C$ is \textit{regular} if, denoted with $\overline{\cal C}$ its closure w.r.t. the topology induced by the inner product of $V$, the conditions $c\in \overline{\cal C}$ and $-c\in \overline{\cal C}$ imply that $c=0$. The intuitive geometrical meaning of this condition is that $\cal C$ is a single cone with a vertex.
\end{itemize}   

It is important to stress that Schrödinger's axioms hold for the so-called \textit{aperture colors} \cite{Hardin:88}, i.e. colored lights seen in isolation against a neutral background or homogeneously colored papers seen through a reduction screen.

Resnikoff dedicated a large part of his paper to motivate the introduction of a new axiom for $\cal C$ and to analyze the strong consequences on its geometry. He postulated that $\cal C$ is a \textit{homogeneous space}, i.e. that there exists a transitive group action on it, which means, in practice, that any couple of elements of $\cal C$ can be connected via an invertible transformation. 

Resnikoff considered this property to be naturally satisfied by the space of perceived colors $\cal C$ because no color can be considered `special' w.r.t. another one. Moreover, he gave an illuminating motivation for his interest in homogeneity by discussing the simplified case of achromatic visual stimuli. These ones provoke only a brightness sensation in humans, for this reason their space can be modeled as $\R^+=(0,+\infty)$ (if we do not consider glare), which is both a group w.r.t. multiplication and a homogeneous space of itself, in fact, for any $x,y\in \R^+$, we can write $y=\lambda x$, with $\lambda=\frac{y}{x}\in \R^+$. Now, the key observation is that, up to a positive constant, the only non-trivial $\R^+$-invariant distance on $\R^+$ is given by 
\begin{equation}
	d(x,y)=|\log(x)-\log(y)|=\left|\log\frac{x}{y}\right|, \qquad x,y\in \R^+,
\end{equation} 
and this expression coincides with the well-known Weber-Fechner's law, the first psycho-physical law ever determined, which establishes the logarithmic response of the human visual system w.r.t. variations of achromatic stimuli, see e.g. \cite{Goldstein:13}. The fact that the only distance compatible with the homogeneous structure of $\R^+$ coincides with a perceptual metric was a major source of inspiration for Resnikoff, who saw in the extension of homogeneity to the 3-dimensional regular convex cone $\cal C$ a possibility to uniquely determine perceptual metrics for the entire color space and not only for the achromatic axis.

\begin{figure}[h!]
	\centering
	\includegraphics[width=2in]{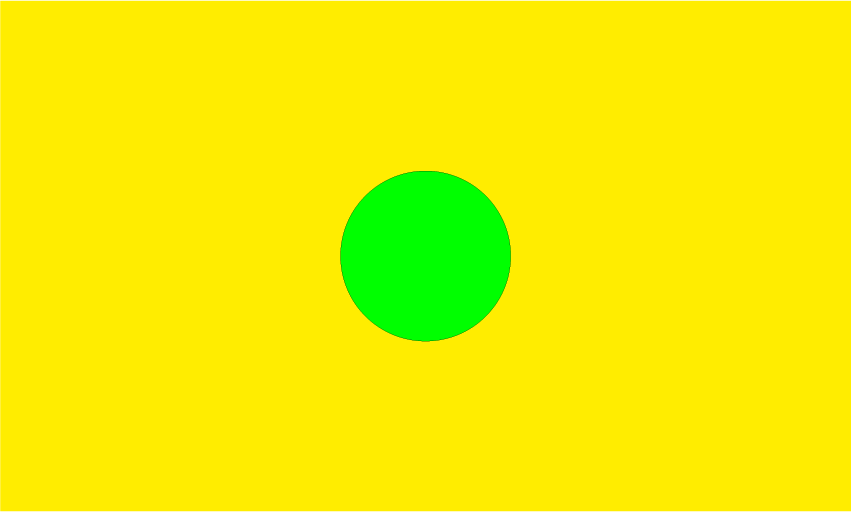}\,
	\includegraphics[width=2in]{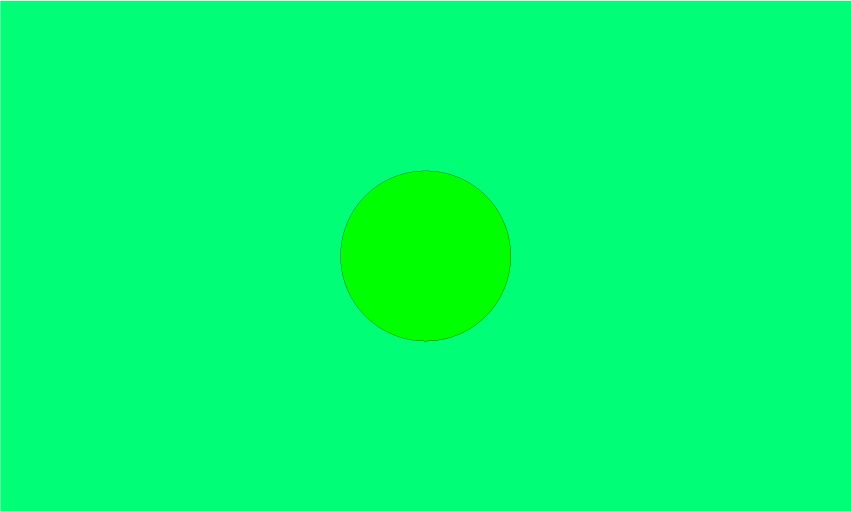}
	\caption{An example background change: in spite of the fact that the inner disks in the center of the two images provide exactly the same physical color stimulus, our perception of them is different because of the so-called chromatic induction phenomenon due to the background difference.}\label{fig:backgroundchange}
\end{figure}

The transitive group action on $\cal C$ described by Resnikoff is that of the so-called `background transformations', operationally implemented by the change of background depicted in Figure \ref{fig:backgroundchange}  and mathematically represented\footnote{The hypothesis of linearity for background transformations remains an open issue, see e.g. \cite{Provenzi:20,Provenzi:20colorationtech}.} by  \textit{orientation-preserving linear transformations that preserve $\cal C$}, i.e.
\beq \text{GL}^+(\mathcal C):=\{B\in \text{GL}(3,\R), \;\text{det}(B)>0 \text{ and } B(c)\in \mathcal C \; \; \forall c\in \cal C\}, \eeq

The hypothesis of homogeneity for $\cal C$ led  to formidable consequences. In fact, Resnikoff proved that there are only two types of 3-dimensional homogeneous regular convex cones:
\beq\label{eq:C1C2}
\begin{cases}
	{\cal C}_1:=\R^+\times\R^+\times\R^+\\
	{\cal C}_2:=\R^+\times \text{SL}(2,\R)/\text{SO}(2)
\end{cases},
\eeq
see also \cite{Provenzi:20} for a simplified proof.

${\cal C}_1=\R^+\times\R^+\times\R^+$ is the classical \textit{flat} colorimetric space used to harbor, e.g., the LMS, RGB, XYZ coordinates \cite{Wyszecky:82}.

Instead, $\text{SL}(2,\R)/\text{SO}(2)$ is a space of constant negative curvature equal to $-1$ and it is an instance of a \textit{2-dimensional hyperbolic model} $\textbf{H}$. Other equivalent models are, e.g., the upper hyperboloid sheet (whose elements can be identified with $2\times 2$ real symmetric positive-definite matrices with unit determinant), the upper half plane and the Poincaré and Klein disks \cite{Ratcliffe:2006}. 

Resnikoff saw in  ${\cal C}_2=\R^+\times \textbf{H}$ a novel way to represent perceived colors: he interpreted $\R^+$ as the brightness axis and $\textbf{H}$ as the chromatic space of perceived colors, thus giving a mathematical formalization to the aforementioned Yilmaz's idea about the pertinence of hyperbolic structures in the study of color perception. 

As we said above, one of Resnikoff's motivations to study a homogeneous color space was the will to uniquely determine perceptual metrics compatible with the homogeneous structure of $\cal C$, i.e. invariant under the action of GL$^+(\cal C)$. He actually succeeded to prove that, for both ${\cal C}={\cal C}_1$ and ${\cal C}={\cal C}_2$, there is only one Riemannian metric $ds^2$, up to positive multiplicative scalars, such that the induced Riemannian distance $d:\cal C \times \cal C \to \R^+$ satisfies
\beq\label{eq:dperceptual}
d(B(c_1),B(c_2))=d(c_1,c_2), \quad \forall c_1,c_2\in \mathcal C, \; \forall B\in \text{GL}^+(\cal C).
\eeq
Specifically, when ${\cal C}={\cal C}_1$, this metric, denoted with $ds_1^2$, is 
\beq \label{eq:metric1}
ds^2= \alpha_1\left(\frac{dx_1}{x_1}\right)^2 + \alpha_2 \left(\frac{dx_2}{x_2}\right)^2 + \alpha_3 \left(\frac{dx_3}{x_3}\right)^2, \qquad x_k,\alpha_k\in \R^+, \; k=1,2,3,
\eeq
which coincides with the well-known Helmholtz-Stiles metric classically used in colorimetry \cite{Wyszecky:82}. Instead, when ${\cal C}={\cal C}_2$ the only  Riemannian metrics satisfying (\ref{eq:dperceptual}) are those positively proportional to 
\beq \label{eq:FisherRao}
ds_2^2 = \frac{1}{2}\text{Tr}((x^{-1} dx)^2), \quad x\in \mathcal H^+(2,\R),
\eeq 
Tr being the matrix trace, which coincides with the Rao-Siegel metric widely used nowadays in geometric science of information, see e.g. \cite{Amari:2012,Calvo:90,Siegel:14}.

\begin{figure}[h!]
	\centering
	\includegraphics[width=5in]{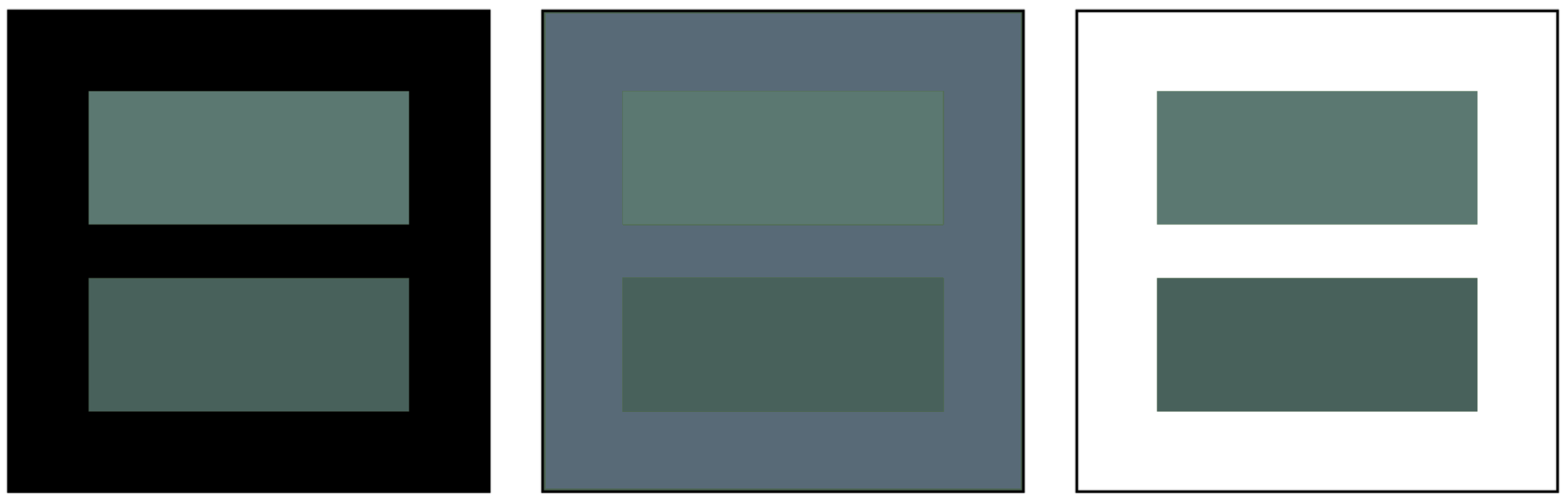}
	\caption{The crispening effect showing that Resnikoff's assumption about the invariance of color metric with respect to background transformations is not coherent with perception.}\label{fig:crispening}
\end{figure}

Assumption (\ref{eq:dperceptual}), however, is not coherent with the so-called \textit{crispening effect} represented in Figure \ref{fig:crispening}, where the same couple of color stimuli is exhibited over three different backgrounds: it is clear that the perceptual difference between them is not background independent.

As a consequence, if background transformations are identified as elements of GL$^+(\mathcal C)$, then neither the Helmholtz-Stiles nor the Rao-Fisher metric can be accepted as perceptually-coherent color distances.

\medskip

In the second part of his 1974 paper, Resnikoff showed how to embed ${\cal C}_1$ and ${\cal C}_2$ in a single mathematical framework thanks to the theory of Jordan algebras. What he lacked to see was the link with a quantum theory of color provided by these objects. We will discuss this in section \ref{sec:quantum}, after recalling the basic results about Jordan algebras and their use in the algebraic formulation of quantum theories.

\section{Jordan algebras and their use in quantum theories}\label{sec:Jalgebras}

Jordan algebras have been introduced by the German theoretical physicist P. Jordan in 1932, see  \cite{Jordan:32}, in the context of quantum mechanics. For the sake of brevity, in this section we will only recap the information about such objects that we need in the sequel, more information can be found in \cite{Faraut:94,Mccrimmon:2004,Baez:12}.

\subsection{Basic results and classification of three-dimensional formally real Jordan algebras and their positive cones}\label{subsec:Jordan}

A Jordan algebra $\mathcal A$ is a real vector space equipped with a bilinear product $(a,b)\mapsto a\circ b$, that is required to be commutative and to satisfy the following Jordan identity:
\begin{equation}
	(a^2\circ b)\circ a=a^2\circ(b\circ a), \qquad \forall a,b\in \cal A,
\end{equation}
which ensures that the power of any element $a$ of $\mathcal A$ is defined without ambiguity and that $\mathcal A$ is at least power-associative, however $\cal A$, in general, is not an associative algebra.  

The most classical example of a non-associative Jordan algebra is given by ${\cal M}(n,\mathbb R)$, the set of real $n\times n$ matrices with $n\ge 2$, equipped with the following \textit{matrix Jordan product}: 
\begin{equation}\label{eq:Jprod}
	X \circ Y = \frac{1}{2}[(X+Y)^2-X^2-Y^2]=\frac{1}{2}(XY+YX), \quad \forall X,Y \in {\cal M}(n,\mathbb R).
\end{equation}
The Jordan algebras that we will consider in the sequel are \textit{formally real}, which means that, for any finite set $a_1,a_2,\ldots,a_n \in \cal A$, $a_1^2+a_2^2+\cdots +a_n^2=0$ implies $a_1=a_2=\cdots=a_n=0$, just as if the elements $a_1,a_2,\ldots,a_n$ were real, which motivates their name. In the sequel, we will make use of the convenient acronym FRJA to denote such Jordan algebras.

It can be proven that any FRJA $\mathcal A$ is unital, i.e. there exists a unit $\textbf{1}\in \cal A$ such that $\textbf{1}\circ a=a \circ \textbf{1}=a$ for all $a\in \cal A$, and it can be endowed with the following partial ordering: for any couple of elements $a,b\in \cal A$, $b\leq a$ if and only if $a-b$ is a sum of squares. In particular, if $a$ is the square of an element of $\cal A$, then we call $a$ \textit{a positive element} and we write $a\geq 0$. The set of positive elements of $\cal A$ is called its \textit{positive domain} and it is denoted with $\overline{\mathcal C}(\mathcal A)$, its interior $\mathcal C(\mathcal A)$ is called the \textit{positive cone} of $\cal A$.

Every FRJA $\mathcal A$ can be equipped with an inner product defined by
\begin{equation}\label{eq:innerproduct}
	\langle a,b\rangle_{\mathcal A} = \text{Tr}(a\circ b), \quad \forall a,b\in \cal A,
\end{equation}
where Tr$(a\circ b)$ actually means Tr$(L_{a\circ b})$, the trace of the endomorphism $L_{a\circ b}:{\cal A}\to {\cal A}$, $c\mapsto (a \circ b) \circ c$, for all $c\in \cal A$. Thus, every FRJA is also a Hilbert space with respect to this inner product.

We now pass to the classification of FRJAs of dimension 3, for more information about a generic dimension $n$ we refer the reader to \cite{Baez:12}.

The classification theorem of Jordan, von Neumann and Wigner \cite{Jordan:34} establishes that there are only two non-isomorphic FRJAs of dimension 3. The first is the \textit{associative} Jordan algebra 
\begin{equation}
	\mathcal A_1=\mathbb R \oplus \mathbb R \oplus \mathbb R
\end{equation}
endowed with the Jordan product $(t_1+t_2+t_3) \circ (s_1+s_2+s_3)=(t_1s_1+t_2s_2+t_3s_3)$, $t_i,s_i\in \R$, $i=1,2,3$. Its positive domain and cone are, respectively, 
\begin{equation}
	\overline{\mathcal C}(\R \oplus \R \oplus \R)=[0,+\infty) \times [0,+\infty) \times [0,+\infty), \quad \mathcal C(\R \oplus \R \oplus \R)=\R^+ \times \R^+ \times \R^+.
\end{equation}
The second option corresponds to two \textit{non-associative} and naturally isomorphic Jordan algebras,  namely:
\begin{equation}
	\mathcal A_2=\mathcal H(2,\mathbb R) \cong \mathbb R \oplus \mathbb R^2,
\end{equation}
where $\mathcal H(2,\mathbb R)$ denotes the Jordan algebra of 2$\times$2 symmetric matrices with real entries equipped with the matrix Jordan product (\ref{eq:Jprod}) and the vector space $\mathbb R \oplus \mathbb R^2$ becomes the so-called \textit{spin factor} when endowed with the Jordan product defined by 
\begin{equation}
	(\alpha+\textbf{v})\circ(\beta+\textbf{w})=((\alpha\beta+\langle \textbf{v},\textbf{w}\rangle) + (\alpha \textbf{w}+\beta \textbf{v})),
\end{equation}
where $\alpha,\beta \in \mathbb R$, $\textbf{v},\textbf{w}\in \mathbb R^2$ and $\langle \, , \, \rangle$ is the Euclidean inner product of $\mathbb R^2$.  The natural isomorphism between the two representations of $\mathcal A_2$ is provided by the following mapping:
\begin{equation}\label{eq:isoH2}
	\begin{array}{cccl}
		\phi: & \mathcal H(2,\mathbb R) & \stackrel{\sim}{\longrightarrow} &  \mathbb R\oplus \mathbb R^2  \\
		& \begin{pmatrix}
			\alpha + v_1 & v_2 \\
			v_2 & \alpha - v_1
		\end{pmatrix}  & \longmapsto         &  (\alpha+\textbf{v}), \qquad \textbf{v}=(v_1,v_2).
	\end{array}
\end{equation}
Thanks to this isomorphism, the positive domains and cones of the two representations of $\mathcal A_2$ are in one-to-one correspondence. Simple  computations show that
\beq\label{eq:cones1}
\begin{array}{ccc}
	\overline{\mathcal C}(\mathbb R \oplus \mathbb R^2) & = & \overline{\mathcal L^+}\\
	\rotatebox{90}{$\cong$} &    & \rotatebox{90}{$\cong$} \\
	\overline{\mathcal C}(\mathcal H(2,\mathbb R)) & = & {\overline{\mathcal H^+}(2,\mathbb R)}   
\end{array}
\eeq 
and
\beq\label{eq:cones2}
\begin{array}{ccc}
	\mathcal C(\mathbb R \oplus \mathbb R^2) & = & \mathcal L^+ \\
	\rotatebox{90}{$\cong$} &    & \rotatebox{90}{$\cong$} \\
	{\mathcal C(\mathcal H(2,\mathbb R))}   & = & {\mathcal H^{+}(2,\mathbb R)}  
\end{array},
\eeq 
where 
\beq
\mathcal L^+=\{(\alpha+\textbf{v})\in \R \oplus \R^2 \; : \; \alpha>0, \; \|(\alpha + \textbf{v})\|^2_{\mathcal M}>0  \},
\eeq 
where $\| \; \|_{\mathcal M}$ is the Minkowski norm defined by $\|(\alpha + \textbf{v})\|^2_{\mathcal M}=\alpha^2 - \|\textbf{v}\|^2$, $\| \; \|$ being the Euclidean norm, is called  \textit{future lightcone} and its closure is
\beq
\overline{\mathcal L^+}=\{(\alpha+\textbf{v})\in \R \oplus \R^2 \; : \; \alpha \ge 0, \; \|(\alpha + \textbf{v})\|^2_{\mathcal M} \ge 0  \},
\eeq 
moreover
\begin{equation}
	\mathcal H^+(2,\mathbb R)=\{X\in \mathcal M(2,\mathbb R) \; : \;  X^t=X \text{ and }\langle \textbf{v},X\textbf{v}\rangle >0 \; \; \forall  \textbf{v}\in \mathbb R^2\setminus \{0\} \},
\end{equation}
is the cone of positive-definite $2\times 2$ real matrices and 
\begin{equation}
	\overline{\mathcal H^{+}}(2,\mathbb R)=\{X\in \mathcal M(2,\mathbb R) \; : \;  X^t=X \text{ and }\langle \textbf{v},X\textbf{v}\rangle \ge 0 \; \; \forall  \textbf{v}\in \mathbb R^2\},
\end{equation}
is the set of positive semi-definite $2\times 2$ real matrices.

For later purposes, we underline here that the positive cone\footnote{Actually $\cal C(A)$ is a \textit{symmetric cone}, i.e. an open convex regular homogeneous self-dual cone, for all FRJA $\cal A$ and, by the Koecher-Vinberg theorem, see e.g. \cite{Koecher:57,Vinberg:61,Faraut:94}, every symmetric cone is isomorphic to the positive cone of a FRJA.} $\cal C(A)$ of a FRJA $\cal A$ has the remarkable property of being \textit{self-dual}, see e.g. \cite{Faraut:94}, i.e. ${\cal C}(\mathcal A)={\cal C}^\ast(\mathcal A)$, where ${\cal C}^\ast(\mathcal A)$ is called dual cone and it is defined as follows
\begin{equation}
	{\cal C}^\ast(\mathcal A) = \{a\in {\cal A} \; : \; \forall b\in {\cal C(A)}, \; \langle a,b\rangle >0\} \cong \{\omega \in {\cal A}^* \; : \; \forall b\in {\cal C(A)}, \; \omega(b) >0\}.
\end{equation}
The isomorphism above is a direct consequence of the Riesz representation theorem which allows us to identify every element $a\in \cal A$ with one and only one linear functional $\omega \in \mathcal A^*$, the dual of the vector space underlying $\cal A$. $\omega\in \mathcal A^*$ is called \textit{positive} if $\omega(a)\ge 0$ for all $a\in \overline{\mathcal C}(\mathcal A)$. If we denote with $\mathcal A^*_+$ the set of positive functionals on $\cal A$ then, by self-duality, we have the identification $\mathcal A^*_+\cong \overline{\mathcal C}(\mathcal A)$, so, thanks to (\ref{eq:cones1}):
\beq
\mathcal H(2,\R)^*_+ \cong \overline{\mathcal H^{+}}(2,\R) \quad \text{ and } \quad (\R \oplus \R^2)^*_+ \cong \overline{\mathcal L^+}.
\eeq 

\medskip
The results that we have recalled so far allow us to show how Resnikoff's finding about the classification of possible perceived color spaces $\mathcal C_1$ and $\mathcal C_2$ appearing in  (\ref{eq:C1C2}) can be related to Jordan algebras theory. In fact, on one side, $\mathcal C_1$ coincides with the positive cone of the associative Jordan algebra $\mathcal A_1=\mathbb R \oplus \mathbb R \oplus \mathbb R$. On the other side, every matrix $X$ of $\mathcal H^+(2,\R)$, the positive cone of the non-associative Jordan algebra $\mathcal A_2=\mathcal H(2,\R)$, has a strictly positive determinant, so we can always decompose it as $X=\sqrt{\det(X)}\frac{X}{\sqrt{\det(X)}}\equiv \sqrt{\det(X)}Y$, where $Y\in \mathcal H^+_1(2,\R)$, the subset of $\mathcal H^+(2,\R)$ given by matrices with unit determinant. Hence $\mathcal H^+(2,\R)\cong \R^+\times \mathcal H^+_1(2,\R) \cong \mathcal C_2$. In other words, the two color space found by Resnikoff can be identified with the positive cones of the only two non-isomorphic formally real Jordan algebras of dimension 3. 

This is as far as Resnikoff went in his paper \cite{Resnikoff:74}. In section \ref{sec:quantum} we will show how to extend Resnikoff ideas to a quantum theory of color perception. The exposition will be clearer if we first explain, in the next subsection, how Jordan algebras relate to quantum theories.

\subsection{Jordan algebras and algebraic formulation of quantum theories}\label{sec:JoQuant}

The birth (and also the name) of quantum mechanics is related to the need to explain the outcomes of physical experiments involving energy quantization. After the early formalization attempts of Born, Heisenberg and Jordan with the so-called `matrix mechanics' and of Schrödinger with his `wave mechanics', first Dirac in \cite{Dirac:82} and then von Neumann in \cite{vonNeumann:2018} provided the abstract setting based on Hilbert spaces and Hermitian operators that, nowadays, we call the ordinary axiomatization of (non-relativistic) quantum mechanics. The work of Dirac and von Neumann is unanimously considered extraordinary not only for their rigorous formalization of quantum mechanics, but also because they boldly gave up preconceptions about nature, such as continuity and deterministic behavior, and simply build the quantum theory from scratch out of the experiments, by searching the minimal mathematical framework and the most suitable names, objects and laws that described the outcome of the experiments. They did not allow previous philosophical dogmas about how nature `should' work to tell them what to do, instead they let nature to speak for itself through mathematics. 

In this sense, quantum mechanics, despite its great mathematical abstractness, is as attached to practical experiments and measurements as it can be. This explains why mathematical definitions always go hand by hand with `operational' definitions in quantum mechanics. Following this tradition, we start by the operational definitions that we will mimic in section \ref{sec:quantum} for a perceptual system:
\begin{itemize}
	\item a \textit{physical system} $\cal S$ is described as a setting where we can perform physical measurements giving rise to quantitative results in conditions that are as isolated as possible from external influences;
	\item a \textit{state} of $\cal S$ is the way it is \textit{prepared} for the measurement of its observables;
	\item \textit{observables} in $\cal S$ are the objects of measures and are associated with the physical apparatus used to measure them on a given state;
	\item an \textit{expectation value} of an observable in a given state is the average result of multiple measurements of the observable when the system is prepared each time in the same state.
\end{itemize}
It is clear that observables characterize a state through their measurements and, vice-versa, the preparation of a particular state characterizes the experimental results that will be obtained on the observables. This \textit{duality observable-state}, as we will see in this subsection, can be formalized mathematically.

In the ordinary mathematical axiomatization of quantum mechanics a physical system is associated to a Hilbert space $\cal H$, a state to a ray of  $\cal H$ (i.e. the linear span of a vector of $\cal H$), an observable to an Hermitian operator $A: \cal H \to \cal H$ and, finally, the expectation value of $A$ on a state is associated to an element of its spectrum. 

Besides this ordinary axiomatization of a quantum system, other, more profound, axiomatizations emerged later. Probably the most general and surely the best suited for our purposes is the so-called \textit{algebraic formulation}, pioneered by Jordan, von Neumann and Wigner in \cite{Jordan:34}. Von Neumann massively contributed to the Hilbert space formalization of quantum mechanics while he was an assistant of Hilbert in Göttingen between 1926 and 1932, year in which he published the book \cite{vonNeumann:2018}. However, he soon came to the conclusion that, from an algebraic point of view, the Hilbert space formulation of quantum mechanics was not optimally suited \cite{Redei:96}: in fact, linear operators on a Hilbert space form an algebra under composition, but two Hermitian operators (associated to quantum observables) are stable under composition if and only if they commute, moreover, on the operational side, the composition of observables makes sense only under restrictive conditions. 

These considerations led him to warmly welcome Jordan's 1932 proposal \cite{Jordan:32} to replace the non-commutative and operationally problematic composition product with the commutative and operationally significant Jordan matrix product defined in eq. (\ref{eq:Jprod}), even if that meant to renounce to associativity. The meaningfulness of the product $a\circ b=\frac{1}{2}(ab+ba)$ lies in Jordan's `brilliantly trivial' observation that it can be re-written as $\frac{1}{2}[(a+b)^2-a^2-b^2]$ or $\frac{1}{4}[(a+b)^2-(a-b)^2]$ where all the operations involved make perfect sense in quantum mechanics, in particular, the square of an observable is simply interpreted as a change of scale in the instrument used for its measurement \cite{Strocchi:08}.

As underlined in \cite{Moretti:17}, the algebraic formulation of a physical theory is perhaps the most general because it can encompass both classical and quantum systems.  The postulates of direct interest for us are the following:
\begin{itemize}
	\item a physical system $\cal S$ is described by its \textit{observables}, which are elements of an algebra $\cal A$ with unit $\textbf{1}$ endowed with a partial ordering. Notice that this does not mean that all the elements of $\cal A$ are observables, but only that the observables of $\cal S$ are contained in $\cal A$;
	\item if $\cal A$ is a commutative and associative algebra, then we deal with a \textit{classical} system; otherwise, we call $\cal S$ a \textit{quantum} system;
	\item a \textit{state} on $\cal A$ is a positive normalized linear functional $\omega: {\cal A} \to \R$, i.e.
	\begin{itemize}
		\item if $a\in \cal A$ is positive, accordingly to the partial ordering of $\cal A$, then $\omega(a)\ge 0$;
		\item $\omega(\textbf{1})=1$.
	\end{itemize}
	\item given an observable $a\in \cal A$ and a preparation of the system, i.e. a state $\omega$, we can associate the number $\langle a\rangle_\omega:=\omega(a)$, called the \textit{expectation value} of the variable $a\in \cal A$ on the state $\omega$. $\omega(a)$ is operationally obtained by performing replicated measurements of $a$ on identically prepared states and by taking the average over the outcomes of measurements.
\end{itemize}
It is important to motivate why the lack of commutativity or associativity of $\cal A$ is the key property to establish the quantum-like character of a theory. The real philosophical and mathematical core of a quantum system is not energy quantization\footnote{in fact, also in quantum mechanics there can be continuous energy bands, e.g. in solid state quantum physics, corresponding to the continuous part of the spectrum of an Hermitian operator.}, but Heisenberg's uncertainty principle \cite{Heisenberg:27}, i.e. paraphrasing the beautiful description contained in \cite{Strocchi:08}, the empirical observation of the existence of observables that cannot be measured simultaneously: the measurement of one of them introduces an unavoidable limit in the precision by which another can be measured, as happens for the observable of the so-called Heisenberg algebra \cite{Levie:20}.

Mathematically, this profound physical fact has nothing to do with the discrete spectrum of an Hermitian operator on a Hilbert space, but with the fact that such operators form an associative but non-commutative algebra. Crucially, Jordan, von Neumann and Wigner proven in \cite{Jordan:34} that also a commutative but non-associative algebra of observables can be used to encode this fundamental fact. Hence, the passage from the associative and commutative algebra of observables (real-valued functions on the phase space) of classical mechanics to the either non-associative or non-commutative algebraic structure of quantum mechanics is the profound and crucial distinction between the two kind of theories.

As we have seen, a FRJA is always commutative, but it can be associative or not, in section \ref{sec:quantum} we will use this property to interpret ordinary colorimetry as a classical theory and to establish a novel quantum theory of color perception through the choice of a non-associative FRJA of observables. A similar use of Jordan algebras has been performed by Emch in statistical mechanics and quantum field theory \cite{Emch:2009}. 

Before doing that, let us complete this section by recalling the fundamental concept of density matrix and its relation to pure and mixed states.

In section \ref{subsec:Jordan} we have seen that Riesz's representation theorem and self-duality imply that  $\mathcal A^*_+ \cong  \overline{\mathcal C}(\mathcal A)$, i.e. the set of positive functionals of a FRJA $\cal A$ can be identified with positive elements of $\cal A$, which form a \textit{closed convex cone}.

Thanks to this result, it is very easy to associate states to elements of $\cal A$ by imposing normalization. For historical reasons, these elements are typically denoted with $\rho$ and called \textit{density matrices}. Recalling eq. (\ref{eq:innerproduct}), the normalization condition can be simply written as follows:
\beq
1=\omega_\rho(\textbf{1})=\langle \rho,\textbf{1}\rangle = \text{Tr}(\rho \circ \textbf{1}) = \text{Tr}(\rho).
\eeq
As a consequence, the `duality state-observable', i.e.  the isomorphism between the set of states $\cal S(\cal A)$ (a subset of $\cal A^*$) and the set of density matrices $\cal DM(A)$ (a subset of $\cal A$), is given by:
\begin{equation}\label{eq:statedensmat}
	\mathcal S(\mathcal A) \cong {\mathcal{DM}}(\mathcal A)=\{\rho \in \overline{\mathcal C}(\mathcal A), \; \text{Tr}(\rho)=1 \}.
\end{equation} 
So, for every state $\omega \in \mathcal S(\mathcal A)\subset \mathcal A^*$, there is one and only one density matrix $\rho \in {\mathcal{DM}}(\mathcal A)\subset \cal A$ such that the expectation value of $a\in \cal A$ is given by:
\beq\label{eq:expval2}
\omega(a) = \langle a \rangle_\omega \equiv  \langle a \rangle_\rho = \text{Tr}(\rho \circ a).
\eeq 
Density matrices can be associated to both pure and mixed states. We say that a state $\omega$ is the mixture of the two states $\omega_1, \omega_2$ if we can write $\omega$ as the convex combination of $\omega_1, \omega_2$, i.e. if there exists $0<\lambda <1$ such that $\omega = \lambda \omega_1 + (1-\lambda)\omega_2$. More generally, a mixed state $\omega$ is the convex combination of $n$ states $\omega_i$, $i=1,\dots,n$, $n\ge 2$, i.e. 
$$\omega = \sum\limits_{i=1}^n \lambda_i \omega_i, \; \text{ with } \sum_{i=1}^n \lambda_i=1.$$ 
A state $\omega$ is \textit{pure} if it cannot be written as a convex linear combination of other states. Geometrically speaking, a pure state does not lie in any open line segment joining two states. So, for example, if the set of states is represented by a disk or a sphere, then the pure states are those lying on the perimeter of the disk or the spherical surface, respectively. In section \ref{sec:quantum} we will confirm this fact for an important case of interest for our theory of color.

Remarkably, if we associate a density matrix $\rho$ to a state $\omega$, then we can establish if $\omega$ is pure or mixed thanks to this very simple criterion, see e.g.  \cite{Moretti:17} or \cite{Blum:2012}:
\begin{itemize} 
	\item $\rho$ describes a pure state if and only if Tr$(\rho^2)=1$;
	\item $\rho$ describes a mixed state if and only if Tr$(\rho^2)<1$.
\end{itemize} 
If we denote with $\mathcal{PS}(\mathcal A)$ the set of pure states of a system with algebra of observables given by the FRJA $\cal A$, then we can write:
\beq 
\mathcal{PS}(\mathcal A)\cong \{\rho \in \overline{\mathcal C}(\mathcal A), \; \text{Tr}(\rho)=1, \; \text{Tr}(\rho^2)=1\} = \{\rho \in \mathcal{DM}(\mathcal A), \; \text{Tr}(\rho^2)=1\}.
\eeq 
Another way to characterize the purity of a state is through the so-called von Neumann entropy, defined as
\begin{equation}\label{eq:vonNentropy1}
	S(\rho)=-\text{Tr}(\rho \circ \log \rho).
\end{equation} 
It is possible to prove that, see e.g. \cite{Heinosaari:2011} or \cite{Petz:2007}:
\begin{itemize} 
	\item if $\cal A$ is a real matrix Jordan algebra, then 
	\beq\label{eq:vonNentropy2}
	S(\rho)=-\sum\limits_{k}\lambda_k \log \lambda_k,
	\eeq 
	where the numbers $\lambda_k$ are the eigenvalues of $\rho$ repeated as many times in the sum as their algebraic multiplicity;
	\item $\rho$ describes a pure state if and only if $S(\rho)=0$ (minimum entropy, i.e. maximal amount of information), if $S(\rho)>0$, $\rho$ describes to a mixed state;
	\item $S$ is invariant under orthogonal conjugation, i.e. $S(O\rho O^t) = S(\rho)$, for all $O\in \text{O}(n)$, $n=\dim(\cal A)$;
	\item $S(\rho)$ is a concave function of $\rho$ in the following sense: $S(t\rho_1+(1-t)\rho_2)\ge tS(\rho_1)+(1-t)S(\rho_2)$ for all couples of density matrices $\rho_1$, $\rho_2$.
	\item the state of maximal entropy, or of minimal amount of information, is the normalized unit of the FRJA $\cal A$:
	\begin{equation}\label{eq:maxentropy}
		\rho_0:=\underset{\rho}{\mathrm{argmax}} \; S(\rho)=\frac{1}{\text{Tr(\textbf{1})}}\textbf{1}\in \cal DM(A).
	\end{equation} 
\end{itemize} 

\section{A quantum theory of color perception}\label{sec:quantum}

We are now ready to state our operational axioms for a theory of color perception by mimicking those of the algebraic formulation of a physical theory:

\begin{itemize}
	\item a \textit{visual scene} is a setting where we can perform psycho-visual measurements in conditions that are as isolated as possible from external influences;
	\item a \textit{perceptual chromatic state} is represented by the preparation of a visual scene for psycho-visual  experiments;
	\item a \textit{perceptual color} is the observable identified with a psycho-visual measurement performed on a given perceptual chromatic state; 
	\item a \textit{perceived color} is the expectation value of a perceptual color after a psycho-visual measurement.
\end{itemize}
It is worthwhile underlying two facts about the previous assumptions:
\begin{itemize}
	\item well-known colorimetric definitions such as additive or substractive synthesis of color stimuli, aperture or surface color, color in context, and so on, are incorporated in the concept of preparation of a perceptual chromatic state. A first example of preparation is the set up a visual scene where an observer in a dark room has to look at a screen, where a light stimulus with foveal aperture w.r.t. the observer provokes a color sensation. A second example of preparation is given by an observer adapted to an illuminant in a room who looks at the patch of a surface. The perceptual chromatic states identified by these two preparations are, in general, different;
	\item the instruments used to measure the observables are not physical devices, but the sensory system of a human being. Moreover, the results may vary from person to person, thus the average procedure needed to experimentally define the expectation value of an observable on a given state is, in general, observer-dependent. The response of an \textit{ideal standard observer} can be obtained through a further statistical average on the observer-dependent expectation values of an observable in a given state.
\end{itemize}
On the mathematical side, the only axiom that we consider is the following, first introduced in \cite{Berthier:2020}:
\smallbreak
{\textsc{Trichromacy axiom}: --} {\em The space of perceptual colors is the positive cone $\mathcal C$ of a formally real Jordan algebra of real dimension 3.} 
\smallbreak
Notice that we associate $\mathcal C$ to perceptual colors, i.e. observable colors, and not to perceived colors, i.e. their expectation values after measurements. In section \ref{sec:discussion} we will motivate the reason why we consider this association more appropriate by discussing the subtle concept of measurement implicitly involved in the definition of perceived color.

As we have discussed in section \ref{subsec:Jordan}, the only formally real Jordan algebra of real dimension 3 are the associative Jordan algebra $\mathbb R \oplus \mathbb R \oplus \mathbb R$ and the non-associative and naturally isomorphic Jordan algebras $\mathcal H(2,\mathbb R) \cong \mathbb R \oplus \mathbb R^2$. The positive cones of $\mathbb R \oplus \mathbb R \oplus \mathbb R$ and $\mathcal H(2,\mathbb R)$ agree exactly with those found by Resnikoff by adding the homogeneity axiom to the set of experimentally well-established Schrödinger's axioms. 

If we transpose the algebraic formulation of physical theories to the case of a perceptual theory of color, we immediately have that:
\begin{itemize}
	\item a theory of color perception associated to the FRJA $\mathbb R \oplus \mathbb R \oplus \mathbb R$ is classical;
	\item a theory of color perception associated to the FRJAs $\mathcal H(2,\mathbb R) \cong \mathbb R \oplus \mathbb R^2$ is quantum-like.
\end{itemize}
As previously seen, standard colorimetry is associated to the FRJA $\mathbb R \oplus \mathbb R \oplus \mathbb R$ and so it is a classical theory associated to a geometrically trivial cone of observables. In \cite{Berthier:2020} the much geometrically richer `quantum colorimetry' associated to the FRJAs $\mathcal H(2,\mathbb R) \cong \mathbb R \oplus \mathbb R^2$ has been investigated by exploiting, among other techniques, the results about density matrices and von Neumann entropy recalled above. 

The results obtained in \cite{Berthier:2020} will be summarized in the following subsections. Before that, we would like to spend a few words to motivate why a quantum theory of color perception is not only a valid option, but, in our opinion, it makes much more sense than a classical one. 

The well-known Copenhagen interpretation of quantum mechanics assumes that the nature of microscopic world, contrary to the macroscopic one, is intrinsically probabilistic and, coherently with that, each quantum measurement must be interpreted in a  probabilistic way. The same elusiveness characterize color perception: it is well-known that the outcome of multiple experiments where an observer must chose, say, a Munsell chip to match a given color patch under a fixed illuminant is, in the large majority of cases, not a sharp choice, but a distribution of close selections around the most frequent one. If we use the terms introduced above we have that, in spite of the fact that the visual scene has been prepared in the same chromatic state, the only way to characterize the perceived color of the patch is through the expectation value of a probability distribution.  

Remarkably, the great theoretical physicist A. Ashtekar and his collaborators, A. Corichi and M. Pierri, foresaw in \cite{Ashtekar:99} the possibility of a quantum theory of color vision by writing: `\textit{the underlying mathematical structure is reminiscent of the structure of states (i.e. density matrices) in quantum mechanics. The space of all states is also convex-linear, the boundary consists of pure states and any mixed state can be obtained by a superposition of pure states. In the present case, the spectral colors are the analogs of pure states}'. In the following subsections we will see how this intuition can be precisely formalized. 

Finally, let us also uphold a peculiar feature of the quantum theory of color perception: it is based on real numbers. This may sound odd, since quantum mechanics is usually thought to be an intrinsically complex theory, this, however, is a misconception. In fact, as remarked in \cite{Beltrametti:2010}, the quantum description of observable as an Hermitian operator acting on a Hilbert space and of state as a density matrix is based upon the spectral theorem and Gleason's theorem, respectively. Both theorems retain their validity on real or quaternionic Hilbert spaces, so, contrary to common belief, a real or quaternionic quantum theory of observables, states and related concepts is as legitimate as a complex quantum theory.

\subsection{Pure and mixed quantum chromatic states}\label{sec:quantstates}
If we specialize eq. (\ref{eq:statedensmat}) in the case $\mathcal A=\mathcal H(2,\R)$ we get:
\begin{equation}
	\mathcal S(\mathcal H(2,\R)) \cong {\mathcal{DM}}(\mathcal H(2,\R))=\{\rho \in \overline{\mathcal C}(\mathcal H(2,\R)), \; \text{Tr}(\rho)=1 \},
\end{equation} 
but, thanks to (\ref{eq:cones1}), $\overline{\mathcal C}(\mathcal H(2,\R))=\overline{\mathcal H^+}(2,\R)$, which is a closed convex cone embedded in a 3-dimensional vector space. The linear constraint Tr$(\rho)=1$ represents a 2-dimensional hyperplane, thus, geometrically, $\mathcal{DM}(\mathcal H(2,\R))$ is expected to be 2-dimensional. If we add the condition Tr$(\rho^2)=1$ to obtain pure states, we further reduce the dimension to 1. Coherently with these considerations, we have:
\beq
\mathcal{S}(\mathcal H(2,\mathbb R))\cong \mathcal{DM}(\mathcal H(2,\mathbb R)) = \left\{ \rho(v_1,v_2)\equiv \frac{1}{2}\begin{pmatrix}
	1+v_1 & v_2 \\ v_2 & 1-v_1
\end{pmatrix}, \text{ with } v_1^2+v_2^2\le 1 \right\}\cong \textbf{\textbf{D}},
\eeq
and	
\beq
\mathcal{PS}(\mathcal H(2,\mathbb R))\cong \left\{ \rho(v_1,v_2)=\frac{1}{2}\begin{pmatrix}
	1+v_1 & v_2 \\ v_2 & 1-v_1
\end{pmatrix}, \text{ with } v_1^2+v_2^2= 1 \right\}\cong S^1,
\eeq
where $\textbf{D}$ is the closed unit disk in $\R^2$ centered in the origin and $S^1$ is its border.

The extension of these results to the spin factor $\R \oplus \R^2$ is obtained immediately by applying the isomorphism defined in (\ref{eq:isoH2}) on the matrices  $\rho(v_1,v_2)$:
$$ \phi(\rho(v_1,v_2))=\phi\left( \frac{1}{2}\begin{pmatrix}
	1+v_1 & v_2 \\ v_2 & 1-v_1
\end{pmatrix} \right) =\frac{1}{2}(1+(v_1,v_2))\equiv \frac{1}{2}(1+\textbf{v})=:s_\textbf{v}$$
so that
\beq\label{eq:statespinfactor}
\mathcal{S}(\R \oplus \R^2)\cong \mathcal{DM}(\R \oplus \R^2) = \left\{ s_{\textbf{v}}= \frac{1}{2}\left(1+\textbf{v} \right), \; \|\textbf{v}\|\le 1 \right\} \cong \textbf{D},
\eeq 
and 
\beq 
\mathcal{PS}(\R \oplus \R^2)\cong \left\{ s_\textbf{v}=\frac{1}{2}\left(1+\textbf{v}\right), \; \|\textbf{v}\|= 1 \right\} \cong S^1.
\eeq
These results show that the states of our quantum color perception theory constitute the real version of the so-called \textit{Bloch sphere}, that, in complex quantum mechanics, harbors the states of a qubit (e.g. an electron with its two spin states), whose Hilbert space state is $\mathbb C^2$. Following Wootters, see e.g. \cite{Wootters:2014}, we call this system a \textit{rebit}, the portmanteau of `real qubit'. 

Since we lose a dimension passing from complex to real numbers, the Bloch sphere of a rebit becomes the Bloch disk $\textbf{D}$: the points in its interior parameterize mixed chromatic states, while those lying on the border $S^1$, parameterize pure chromatic states.

\subsection{Von Neumann entropy of quantum chromatic states: saturation and hue}\label{sec:vonNeumannsat}

Thanks to the von Neumann entropy, we can provide an explicit measure of the degree of purity of a quantum chromatic state. Let us start with the maximal von Neumann entropy, expressed by formula (\ref{eq:maxentropy}): the identity element of $\mathcal A=\mathcal H(2,\mathbb R)$ is $I_2=2\rho(0,0)$ and Tr$(I_2)=2$, so 
$$ \rho_0:={1 \over 2} I_2 =\rho(0,0)\in \mathcal H(2,\R), $$
i.e. the state of maximal von Neumann entropy for $\mathcal H(2,\R)$ is parameterized by the origin of the unit disk $\bf D$.

By using the isomorphism defined in (\ref{eq:isoH2}), we can obtain the density matrix of associated to the maximal von Neumann entropy in the case $\mathcal A=\R \oplus \R^2$, obtaining:
\beq
\phi(\rho_0)=\frac{1}{2}(1+\textbf{0}) \in \R \oplus \R^2.
\eeq
The von Neumann entropy of a generic quantum chromatic state can be computed easily if we express the parameters $(v_1,v_2)\in \textbf{D}$ of a density matrix to the polar form $(r\cos \vartheta,r\sin \vartheta)$, with $r\in [0,1], \; \vartheta\in [0,2\pi)$, the most natural parameterization of the disk $\textbf{D}$, obtaining:
\begin{equation}\label{eq:rhortheta}
	\rho(r,\vartheta)=\frac{1}{2}\begin{pmatrix}
		1+r\cos\vartheta&r\sin\vartheta\\r\sin\vartheta & 1-r\cos\vartheta \end{pmatrix}.
\end{equation}
Thanks to formula (\ref{eq:vonNentropy2}), by direct computation we get that the von Neumann Entropy of a quantum state described by the density matrix $\rho(r,\vartheta)$ can be written as follows:
\beq
S(\rho(r,\vartheta))=S(r)=-\log \Big( \frac{1-r}{2}\Big)^{\frac{1-r}{2}}\Big(\frac{1+r}{2}\Big)^{\frac{1+r}{2}}=\log2 -\log(1-r) -(1+r)\tanh^{-1}(r) \, , \; 0\le r<1,
\eeq 
and $S(1):=\lim\limits_{r\to 1^{-}}S(r)=0$. This explicit expression is coherent with the properties of von Neumann entropy listed in section \ref{sec:JoQuant}: $S(\rho(r,\vartheta))$ is a radial concave bijective function on $[0,1]$, its maximum and minimum values are $\log(2)$ and 0, reached in correspondence of $r=0$ and $r=1$, respectively. 

Since $r=0$ identifies the maximal von Neumann entropy of a quantum chromatic state and $r=1$ identifies pure quantum chromatic states, it seems reasonable to associate the von Neumann entropy to the saturation of a perceptual color: when $r=0$ we have the minimal chromatic information available, i.e. $r=0$ describes achromatic colors; when $r=1$, instead, we have the maximal chromatic information, i.e. $r=1$ describes fully saturated colors and this holds for all values of the angle $\vartheta\in [0,2\pi)$. This implied that it also seems reasonable to associate the density matrices
\begin{equation}\label{eq:rhortheta}
	\rho(1,\vartheta)=\frac{1}{2}\begin{pmatrix}
		1+\cos\vartheta&\sin\vartheta\\\sin\vartheta & 1-\cos\vartheta \end{pmatrix}, \quad \vartheta\in [0,2\pi),
\end{equation}
to quantum chromatic states of `pure hue'. 

\begin{figure}[!ht]
	\begin{center}
		\includegraphics[width=7cm]{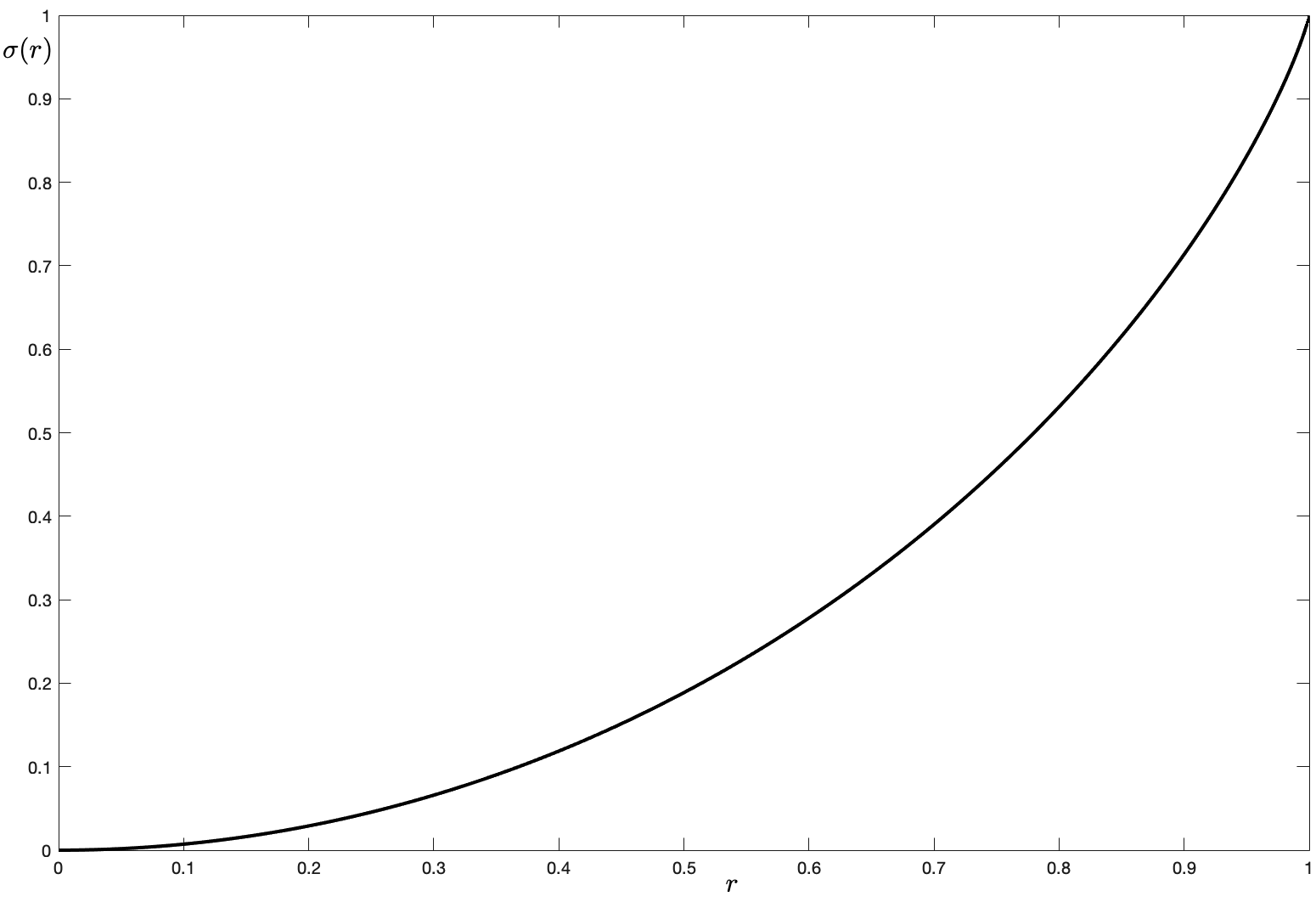}
		\caption{A proposal for the saturation of a quantum chromatic state from built from its von Neumann entropy.}
		\label{fig:saturation}
	\end{center}
\end{figure}

The easiest way to obtain a saturation formula from the von Neumann entropy that associates to achromatic colors the value 0 and to pure hues the value 1 is the following:
\beq
\sigma(r):=\frac{\log(2) - S(r)}{\log(2)}=\frac{\log(1-r) + (1+r)\tanh^{-1}(r)}{\log(2)}=\log_2(1-r)+\frac{1}{2}(1+r)\log_2\left(\frac{1+r}{1-r}\right).
\eeq 
The graph of $\sigma(r)$ can be seen in Figure \ref{fig:saturation}. Its convex behavior, with a small slope near $r=0$, a linear behavior near $r=1/2$ and a large slope near $r=1$ seems to fit well with common perception. The definition of saturation is probably the most elusive among color attributes, thus we are very interested to conduct careful tests in collaboration with psycho-physicists in order to validate or improve this definition of saturation, recalling once more that we are modeling color perception in very restrictive conditions and not in natural scenes.

\subsection{Hering's chromatic opponency and its role in the encoding of visual signals}\label{sec:Hering}

One of the most important results of \cite{Berthier:2020} is a mathematical explanation of Hering opponency \cite{Hering:1878}, i.e. the fact that no color is simultaneously perceived as reddish and greenish, or as yellowish and bluish and, even more importantly, that the encoding of light signals performed by the human visual system, i.e. the superposition of achromatic plus chromatically opponent information performed mainly by ganglion cells, see e.g. \cite{Hubel:95}, can be intrinsically described by the mathematical framework of the quantum theory of color perception. This is very important because, up to the authors' knowledge, only an \textit{a posteriori} explanation of this physiological behavior based on natural image statistics was available, see e.g. \cite{Buchsbaum:83,Ruderman:98,Provenzi:16} and the references therein.

In order to obtain the results claimed above it is necessary to introduce the two Pauli-like matrices $\sigma_1$, $\sigma_2$ given by
\begin{equation}
	\sigma_1=\left(\begin{array}{cc}1 & 0 \\0 & -1\end{array}\right),\ \ \sigma_2=\left(\begin{array}{cc}0 & 1 \\1 & 0\end{array}\right),
\end{equation}
and to notice that the set $(\sigma_0,\sigma_1,\sigma_2)$, where $\sigma_0=I_2$, is a basis for $\mathcal H(2,\R)$.

By direct computation, the generic density matrix of $\mathcal H(2,\R)$ can be obtained from this basis as follows:
$$
\rho(v_1,v_2)= {1\over 2}(I_2+v_1\sigma_1 + v_2\sigma_2)=\rho_0 +{1\over 2}(v_1\sigma_1 + v_2\sigma_2) , \qquad \|(v_1,v_2)\|\le 1,
$$
or, in polar coordinates,
\beq\label{eq:rhogeneric}
\rho(r,\vartheta)= \rho_0 +\frac{r\cos \vartheta}{2}\sigma_1 + \frac{r\sin \vartheta}{2}\sigma_2, \qquad r\in [0,1], \; \vartheta \in [0,2\pi).
\eeq 
Let us express $\sigma_1$ and $\sigma_2$ in terms of suitable density matrices by considering the following pure state density matrices corresponding to noticeable values of the angle $\vartheta$:
\beq\label{eq:specialangles}
\rho(1,0)=\rho_0+\frac{\sigma_1}{2} , \quad
\rho(1,\pi)=\rho_0-\frac{\sigma_1}{2}, \quad
\rho\left(1,{\pi\over 2}\right)=\rho_0+\frac{\sigma_2}{2}, \quad 
\rho\left(1,{3\pi\over 2}\right)=\rho_0-\frac{\sigma_2}{2},
\eeq 
we have: 
\beq\label{eq:sigma1}
\sigma_1=\rho(1,0)-\rho(1,\pi) \quad \text{ and } \quad \sigma_2=\rho\left(1,{\pi\over 2}\right)-\rho\left(1,{3\pi\over 2}\right),
\eeq 
by introducing these expressions in eq. (\ref{eq:rhogeneric}) we arrive at the formula
\begin{equation}
	\rho(r,\vartheta)=\rho_0+{r\cos\vartheta\over 2}\left[\rho(1,0)-\rho(1,\pi)\right]+{r\sin\vartheta\over 2}\left[\rho\left(1,{\pi\over 2}\right)-\rho\left(1,{3\pi\over 2}\right)\right] \, ,
\end{equation}
which implies that the generic quantum chromatic state represented by a density matrix $\rho(r,\vartheta)$, with $(r\cos \vartheta,r\sin \vartheta)\in \textbf{D}$ can be seen as the superposition of:

\begin{itemize} 
	\item the maximally von Neumann entropy state $\rho_0$, which represents the achromatic state;
	\item two pairs of diametrically opposed pure hues.
\end{itemize}
This purely theoretical result can be connected to Hering's theory \cite{Hering:1878} by identifying the pairs of pure hues with red vs. green and yellow vs. blue, or to the neural coding theory of de Valois \cite{Devalois:97} by identifying them with pinkish-red vs. cyan and violet vs. greenish-yellow. In spite of the particular identification, the important fact to underline is that our framework allows to intrinsically represent the retinal encoding of visual signals without the need of any \textit{a posteriori} analysis or manipulation. 

We note also that if we sum all the density matrices listed in eq. (\ref{eq:specialangles}) we get $4\rho_0$, so 
\begin{equation}
	\rho_0 ={1\over 4}\rho(1,0)+{1\over 4}\rho(1,\pi)+{1\over 4}\rho\left(1,{\pi\over 2}\right)+{1\over 4}\rho\left(1,{3\pi\over 2}\right),
\end{equation}
i.e. the achromatic state $\rho_0$ is the mixed state obtained by a convex combination of pure chromatic states where each one of them appears with the same probability coefficient. This fact further supports the interpretation of $\rho_0$ as the achromatic state.

The expectation values of the Pauli-like matrices $\sigma_1$ and $\sigma_2$ on the chromatic state represented by $\rho(r,\vartheta)$ carry a very important information. To compute them, we can use the formula
\beq
\langle \sigma_i \rangle_{\rho(r,\vartheta)} =\text{Tr}(\rho(r,\vartheta) \circ \sigma_i), \qquad i=1,2,
\eeq
which, by direct computation, gives
\beq\label{eq:expsigmas}
\langle \sigma_1 \rangle_{\rho(r,\vartheta)}=r\cos \vartheta \quad \text{ and } \quad  \langle \sigma_2 \rangle_{\rho(r,\vartheta)} = r\sin \vartheta.
\eeq 
Being the cosine and the sine the projection of the unit vector in the disk $\textbf{D}$ on the horizontal and vertical axis, respectively, the interpretation of the previous results is immediate: \textit{the expectation value of $\sigma_1$ (resp. $\sigma_2$) is the degree of opposition between the two pure color states that lie at extreme points of the horizontal (resp. vertical) segment $[-1,1]$}.

\begin{figure}[!ht]
	\begin{center}
		\includegraphics[width=7cm]{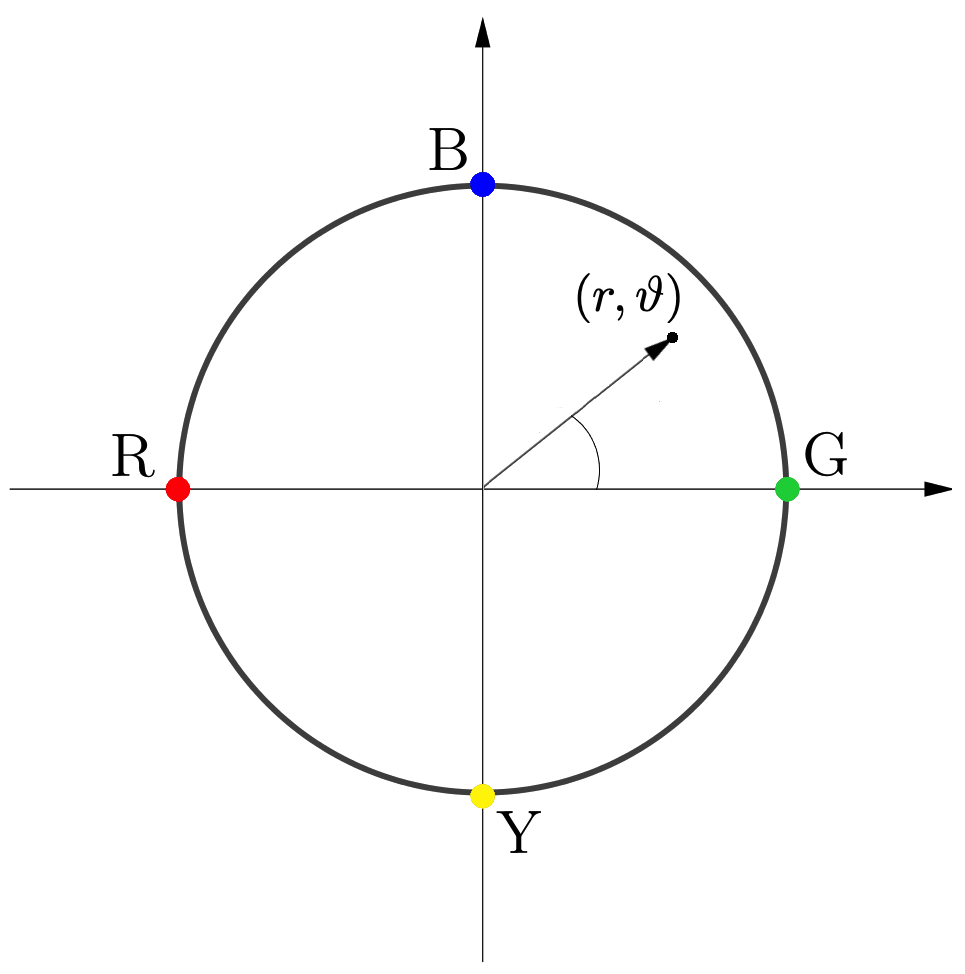}
		\caption{The unit (Bloch) disk $\textbf{D}$ in $\R^2$, whose points represent quantum chromatic states. The points of its  border $S^1$ represent pure quantum chromatic states and the center represents the achromatic state.}
		\label{fig:rebit}
	\end{center}
\end{figure}

Figure \ref{fig:rebit} gives a graphical representation of what was just stated. It can be interpreted as a mathematically rigorous quantum version of Newton's circle. 

In the model that we have described, Hering's observation about the fact that a color cannot be perceived simultaneously as greenish and reddish, or as bluish and yellowish, is an immediate consequence of the fact that the value of $\cos \vartheta$ and $\sin \vartheta$ cannot be simultaneously positive and negative. 

Each pure state associated to a hue can be represented by the density matrix 
$$\rho(1,\vartheta)=\frac{1}{2} \begin{pmatrix}
	1+\cos \vartheta & \sin \vartheta \\
	\sin \vartheta & 1- \cos \vartheta
\end{pmatrix}=\frac{1}{2} \begin{pmatrix}
	1+\langle \sigma_1 \rangle_{\rho(1,\vartheta)} & \langle \sigma_2 \rangle_{\rho(1,\vartheta)} \\
	\langle \sigma_2 \rangle_{\rho(1,\vartheta)} & 1- \langle \sigma_1 \rangle_{\rho(1,\vartheta)}
\end{pmatrix},$$
or, equivalently, to the point $(\cos \vartheta,\sin \vartheta)\in S^1$. This implies that, in this model of color perception, each hue is uniquely associated to a pair numbers belonging to $[-1,1]$ representing the degrees of opposition red-green and blue-yellow of that particular hue. 

\subsection{Uncertainty relations for chromatic opponency}\label{sec:uncertainty}

As we have previously said, the essence of a quantum theory is the existence of uncertainty relations for some observables. In this subsection we are going to derive such uncertainty relations for a generic couple of observables evaluated on arbitrary states and then we will apply our result on the Pauli-lile matrices to show uncertainty in the measurement of the degrees of chromatic opponency described in the previous subsection. 

In order to obtain these relations, we will not make use of the original Heisenberg's or Weyl's arguments, see \cite{Heisenberg:27} and  \cite{Weyl:27} respectively, instead, we will use Robertson's and, in particular, Schrödinger's refinements, see \cite{Robertson:29} and \cite{Schroedinger:30} respectively.

Given an element $a\in \mathcal H(2,\R)$ and a density matrix $\rho \in \mathcal{DM}(\mathcal H(2,\R))$, the quadratic dispersion (or variance) of $a$ on the state $\rho$ is:
\beq (\Delta a_\rho)^2:=\langle a^2 \rangle_\rho - (\langle a \rangle_\rho)^2 = \text{Tr}(\rho \circ a^2)-(\text{Tr}(\rho \circ a))^2. 
\eeq 
For the sake of a simpler proof of our result, let us first define
\beq
\tilde a_\rho := a-\langle a \rangle_\rho I_2= a - \text{Tr}(\rho \circ a)I_2,
\eeq 
and observe that
\beq\label{eq:useful01}
\begin{split}
	\text{Tr}(\rho \circ \tilde a_\rho^2) &  = \text{Tr}(\rho \circ a^2) -2 \text{Tr}(\rho \circ a) \text{Tr}(\rho \circ a) + (\text{Tr}(\rho \circ a))^2 \text{Tr}(\rho) \\
	& =  \text{Tr}(\rho \circ a^2) - (\text{Tr}(\rho \circ a))^2\\
	& = (\Delta a_\rho)^2.
\end{split}
\eeq 

\noindent \textbf{Proposition}:
For all $a,b\in \mathcal H(2,\R)$ and for all density matrices $\rho \in \mathcal{DM}(\mathcal H(2,\R))$ it holds that 
\beq\label{eq:SchrJordan}
(\Delta a_\rho)^2(\Delta b_\rho)^2 \geq \left [ \operatorname{Tr}(\rho \circ (a\circ b)) - \operatorname{Tr}(\rho \circ a)\operatorname{Tr}(\rho \circ b)\right ]^2.
\eeq 

\proof
For all $\gamma\in \R$, it will prove to be useful to analyze the following quantity 
\beq
\text{Tr}(\rho \circ (\tilde a_\rho + \gamma \tilde b_\rho)^2) = \text{Tr}(\rho \circ \tilde a_\rho^2) + 2\gamma \text{Tr}(\rho \circ (\tilde a_\rho \circ \tilde b_\rho)) + \gamma^2 \text{Tr}(\rho \circ \tilde b_\rho^2)). 
\eeq 
Of course, $\text{Tr}(\rho \circ (\tilde a_\rho + \gamma \tilde b_\rho)^2)\ge 0$ for all $\gamma\in \R$. If we fix $a,b$ and $\rho$, then we can consider the function $\varphi:\R \to [0,+\infty)$, $\gamma\mapsto \varphi(\gamma):=\text{Tr}(\rho \circ (\tilde a_\rho + \gamma \tilde b_\rho)^2)$ and search for  
\beq 
\gamma^*:=\underset{\gamma\in \mathbb R}{\text{argmin}} \; \varphi(\gamma).
\eeq 
From $\varphi'(\gamma)=2 \text{Tr}(\rho \circ (\tilde a_\rho \circ \tilde b_\rho)) + 2\gamma \text{Tr}(\rho \circ \tilde b_\rho^2))$ we obtain 
\beq
\gamma^*=- \frac{\text{Tr}(\rho \circ (\tilde a_\rho \circ \tilde b_\rho))}{\text{Tr}(\rho \circ \tilde b_\rho^2)},
\eeq 
so $\varphi(\gamma^*)$, the minimum value attained by $\text{Tr}(\rho \circ (\tilde a_\rho + \gamma \tilde b_\rho)^2)$ as $\gamma$ varies in $\R$, is
\beq
\min_{\gamma\in \mathbb R} \; \varphi(\gamma)= \varphi(\gamma^*)=\text{Tr}(\rho \circ \tilde a_\rho^2) - \frac{\left[\text{Tr}(\rho \circ (\tilde a_\rho \circ \tilde b_\rho))\right]^2}{\text{Tr}(\rho \circ \tilde b_\rho^2)} \ge 0,
\eeq 
which implies
\beq
\text{Tr}(\rho \circ \tilde a_\rho^2) \text{Tr}(\rho \circ \tilde b_\rho^2) \ge \left[\text{Tr}(\rho \circ (\tilde a_\rho \circ \tilde b_\rho))\right]^2,
\eeq 
or, thanks to (\ref{eq:useful01}), 
\beq
(\Delta a_\rho)^2(\Delta b_\rho)^2 \ge \left[\text{Tr}(\rho \circ (\tilde a_\rho \circ \tilde b_\rho))\right]^2.
\eeq 
To conclude the proof, we just have to make the right-hand side of the previous inequality explicit:
\beq
\begin{split}
	\text{Tr}(\rho \circ (\tilde a_\rho \circ \tilde b_\rho)) & = \text{Tr}(\rho \circ ((a-\text{Tr}(\rho \circ a)I_2) \circ (b-\text{Tr}(\rho \circ b)I_2)))\\
	& = \text{Tr}(\rho \circ (a\circ b -\text{Tr}(\rho \circ a) b - \text{Tr}(\rho \circ b) a+\text{Tr}(\rho \circ a) \text{Tr}(\rho \circ b)I_2)) \\
	& = \text{Tr}(\rho \circ (a\circ b)) - \text{Tr}(\rho \circ a)\text{Tr}(\rho \circ b),
\end{split}
\eeq 
i.e. 
\beq 
(\Delta a_\rho)^2(\Delta b_\rho)^2 \ge \left[\text{Tr}(\rho \circ (a\circ b)) - \text{Tr}(\rho \circ a)\text{Tr}(\rho \circ b)\right]^2.
\eeq 
\qed 

\begin{figure}[!ht]
	\begin{center}
		\includegraphics[width=10cm]{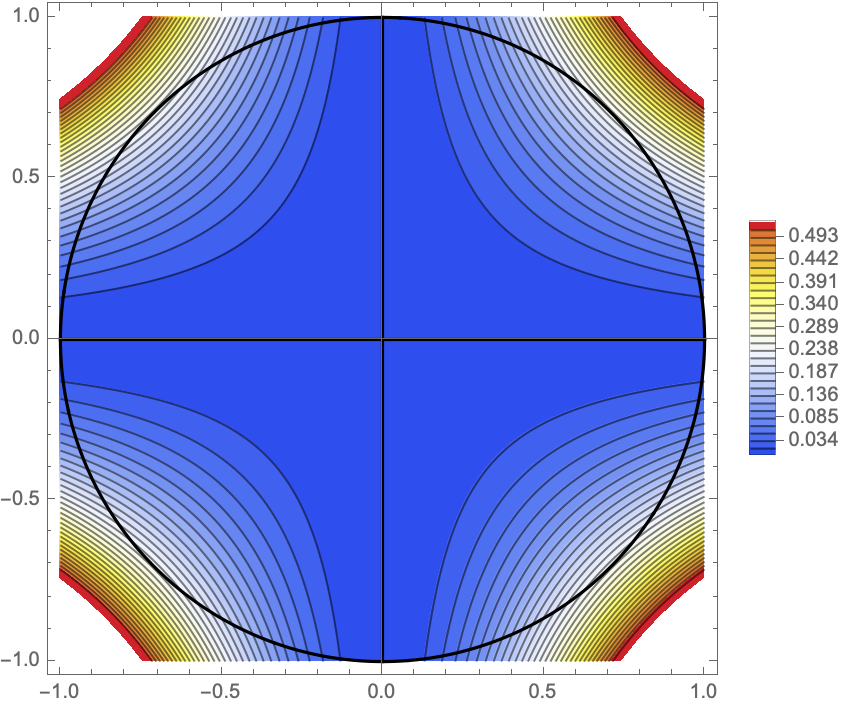}
		\caption{The polar graph of the lower bound $\frac{r^4}{4} \sin^2(2\vartheta)$ of eq. (\ref{eq:SchrPauli}). Notice that this lower bound is 0 only for points with coordinates $(r,\vartheta)$ lying on the opponency axes, identified with the horizontal and vertical axes in the picture. The lower bound is maximal for $(r,\vartheta) = (1,\frac{\pi}{4}+k\frac{\pi}{2})$, $k=1,2,3$.}
		\label{fig:uncertainty}
	\end{center}
\end{figure}

Let us apply the uncertainty relation just proven by identifying $a$ and $b$ with the Pauli-like matrices $\sigma_1$ and $\sigma_2$, respectively, and by using the polar expression $\rho(r,\vartheta)$ for the density matrices. By noticing that $\sigma_1 \circ \sigma_2 =[\sigma_1 \sigma_2 + \sigma_2\sigma_1]/2 =  \textbf{0}$, we get that Tr$(\rho \circ (\sigma_1 \circ \sigma_2))=0$, thus, by using the expectation values computed in (\ref{eq:expsigmas}), we obtain 
\beq\label{eq:SchrPauli}
(\Delta {\sigma_1}_\rho)^2(\Delta {\sigma_2}_\rho)^2 \geq r^4 \cos^2(\vartheta) \sin^2(\vartheta)= \frac{r^4}{4} \sin^2(2\vartheta),
\eeq 
for all $r\in [0,1]$ and $\vartheta \in [0,2\pi)$. The graph of $\frac{r^4}{4} \sin^2(2\vartheta)$ in polar coordinates is depicted in Figure \ref{fig:uncertainty}.

If we identify the opponent axes with red vs. green (R-G) and yellow vs. blue (Y-B), then the interpretation of formula (\ref{eq:SchrPauli}) is the following:
\begin{itemize}
	\item the only value of $r$ that nullifies the right hand side of (\ref{eq:SchrPauli}) is $r=0$, that corresponds to an achromatic color state, thus  \textit{no uncertainty about the degree of opposition R-G and Y-B is present in this case}. This fact is coherent with both our physiological and perceptual knowledge of color perception: a perceived achromatic color is characterized by an `equal amount of chromatic opponencies';
	\item the only values of $\vartheta$ that nullify the right hand side of (\ref{eq:SchrPauli}) are $\vartheta=0,\frac{\pi}{2},\pi,\frac{3}{2}\pi$, which identify the two opposition axes R-G and Y-B. Again, this seems coherent with common knowledge: suppose that we want to determine the couple of opponencies R-G and Y-B to match, say, a color perceived as a red but with a non maximal saturation (measured as a function of the von Neumann entropy of its chromatic state). Due to the redness of the percept, we will always set an equal opposition in the axis Y-B that will not influence the search for the correct opposition R-G. Thus, the determination of the two chromatic oppositions will be compatible;
	\item instead, for $r\in (0,1]$ and $\vartheta \in [0,2\pi)\setminus \{0,\frac{\pi}{2},\pi,\frac{3}{2}\pi\}$, there will be a lower bound strictly greater than 0 for the product of quadratic dispersions of $\sigma_1$ and $\sigma_2$ on the state defined by $\rho(r,\vartheta)$. Moreover, this lower bound is a non-linear function of the variables $(r,\vartheta)$ and it is maximum for pure hues, $r=1$, halfway in between G and B, B and R, R and Y and Y and G, i.e. $\vartheta = \frac{\pi}{4}+k\frac{\pi}{2}$, $k=1,2,3$. If this interpretation is correct, then trying to adjust the R-G opposition to match, say, a color perceived as orange, should introduce a `perceptual disturbance' on the adjustment of the opposition Y-B.
\end{itemize}

Even if the implications of the uncertainty relations discussed above seem coherent with common perception, they remain purely theoretical at the moment, thus they need to be validated by accurate experiments. If the validation will turn out to be faithful with the predictions of formula (\ref{eq:SchrPauli}), then this will provide a further firm confirmation of the non-classical nature of color perception.

\subsection{Geometry and metrics of quantum chromatic states}\label{sec:geometryofQstates}

In this section we will deal with the geometry and metric of the perceived color space and of quantum chromatic states following \cite{Berthier:2020}.

We start by noticing that every matrix belonging to
$\mathcal H^+_{1}(2,\R)=\{X\in\mathcal H^+(2,\mathbb R), \;\det(X)=1\}$
can be written as
\begin{equation}
	X=\left(\begin{array}{cc}\alpha+v_1 & v_2 \\ v_2 & \alpha-v_1\end{array}\right)\ ,
\end{equation}
with $\alpha>0$, to guarantee the positive-definiteness, and $v=(v_1,v_2) \in \mathbb R^2$ satisfying $\alpha^2-\Vert v\Vert^2=1$ to guarantee that $\det(X)=1$. Thanks to the isomorphism (\ref{eq:isoH2}), $\mathcal H^+_{1}(2,\R)$ is in one-to-one correspondence with the level set of the future lightcone $\cal L^+$ given by 
\begin{equation}
	\mathcal L_1=\{(\alpha+\textbf{v})\in\mathcal L^+,\ \|(\alpha +\textbf{v})\|^2_{\mathcal M}=1\}.
\end{equation} 
As proven for instance in \cite {Cannon:97}, the projection on the plane in $\R \oplus \R^2$ identified by the condition $\alpha =0$, i.e.
\begin{equation}\label{eq:Poincare}
	\begin{array}{cccl}
		\pi_1: & \mathcal L_1 & \longrightarrow & \{\alpha=0\}  \\
		& (\alpha+\textbf{v}) & \longmapsto         & \left(0+\frac{\textbf{v}}{1+\alpha}\right)\equiv (0+\textbf{w})
	\end{array}
\end{equation}
is an isometry between $\mathcal L_1$ and the \textit{Poincaré disk} $\cal D$. 

The expression of the matrix $X$ in the $\textbf{w}$-parametrization, with of course $w_i=\frac{v_i}{1+\alpha}$, $i=1,2$, is
\begin{equation}
	\label{eq:paraw}
	X=\left(\begin{array}{cc}\displaystyle {1+2w_1+(w_1^2+w_2^2)\over 1-(w_1^2+w_2^2)} & \displaystyle {2w_2\over 1-(w_1^2+w_2^2)} \\ \displaystyle {2w_2\over 1-(w_1^2+w_2^2)} & \displaystyle {1-2w_1+(w_1^2+w_2^2)\over 1-(w_1^2+w_2^2)}\end{array}\right).
\end{equation}
and, when written like that, every $X\in \mathcal H^+_{1}(2,\R)$ satisfies the following equation, where $ds^2_{\cal D}$ represents the Riemannian metric of the Poincaré disk \cite{Berthier:2020}:
\begin{equation}\label{eq:RSP}
	ds^2_{\cal{D}}=4 \, {(dw_1)^2+(dw_2)^2\over (1-(w_1^2+w_2^2))^2}=\frac{1}{2}{\rm Tr}\left[(X^{-1}dX)^2\right].
\end{equation} 

Thanks to the decomposition $\mathcal H^+(2,\R)=\R^+\times \mathcal H^+_{1}(2,\R)$, we have that $\mathcal H^+(2,\R)$, the positive cone of $\mathcal H(2,\R)$, is foliated with leaves isometric to the Poincaré disk. We recall that the left hand side of eq. (\ref{eq:RSP}) is the Rao-Siegel metric encountered in section \ref{sec:Resnikoff} when we have discussed the Resnikoff model, with the difference that Resnikoff applied it on the whole cone $\mathcal H^+(2,\R)$ and not on the level set $\mathcal H^+_{1}(2,\R)$. This metric has been used also in \cite{Lenz:99} and in \cite{Chevallier:18} in the context of CIE (Commission Internationale de l'Éclairage) colorimetry.

This description is not the best suited for the metric purposes of the quantum theory of color perception because it does not take into account the role of density matrices. We are going to show that the correct way to deal with this issue is by considering an alternative description based on the spin factor $\R \oplus \R^2$ and its positive cone $\cal L^+$. 

Again in \cite{Cannon:97} we can find a classic result of hyperbolic geometry which says that the projection:
\begin{equation}\label{eq:Klein}
	\begin{array}{cccl}
		\tilde \pi_1 : & \mathcal L_1 & \longrightarrow & \{\alpha=1\}  \\
		&  (\alpha+\textbf{v})& \longmapsto         &\left(1+\frac{\textbf{v}}{\alpha}\right)\equiv (1+\textbf{x}),
	\end{array}
\end{equation}
with $x_i=\frac{v_i}{\alpha}$, $i=1,2$, is an isometry between the level set $\mathcal L_1$ and the \textit{Klein disk} $\mathcal K$, whose Riemannian metric is given by:
\begin{equation}
	ds^2_{\mathcal K}={(dx_1)^2+(dx_2)^2\over 1-(x_1^2+x_2^2)}+{(x_1dx_1+x_2dx_2)^2\over (1-(x_1^2+x_2^2))^2}.
\end{equation}
Moreover, the Klein and Poincaré disks, $\mathcal K$ and $\cal D$, are isometric via the map defined by:
\begin{equation}\label{eq:KD}
	x_i={2w_i\over 1+(w_1^2+w_2^2)}, \quad w_i={x_i\over 1+\sqrt{1-(x_1^2+x_2^2)}}, \quad i=1,2. 
\end{equation}
It is possible to verify that $\mathcal L^+$, the positive cone of $\R \oplus \R^2$,  is foliated by the level sets $\alpha$=constant $>0$ with leaves isometric to the Klein disk $\mathcal K$. 

The leaf associated to $\alpha=1/2$ is particularly important. The reason is easy to understand: we have seen in (\ref{eq:statespinfactor}) that the state density matrices $\rho(v_1,v_2)$ can be identified with the elements $s_\textbf{v}=\frac{1}{2}(1+\textbf{v})=\left(\frac{1}{2}+\frac{\textbf{v}}{2}\right)$ with $\Vert \textbf{v}\Vert\leq 1$ of the spin factor $\mathbb R\oplus\mathbb R^2$. 

If we set 
\beq 
\mathcal L_{1/2}=\left\{\left(\frac{\alpha}{2}+\frac{\textbf{v}}{2}\right)\in \mathcal L^+, \;  \left\|\left(\frac{\alpha}{2}+\frac{\textbf{v}}{2}\right)\right\|^2_{\cal M}=\frac{1}{4}\right\},
\eeq 
then the projection:
\begin{equation}\label{eq:Kleinhalf}
	\begin{array}{cccl}
		\tilde \pi_{1/2}: & \mathcal L_{1/2} & \longrightarrow & \{\alpha=1/2\}  \\
		& \left(\frac{\alpha}{2}+\frac{\textbf{v}}{2}\right)  & \longmapsto         & \left(\frac{1}{2}+\frac{\textbf{v}}{2\alpha}\right),
	\end{array}
\end{equation}
satisfies $\tilde \pi_{1/2}\left(\frac{\alpha}{2}+\frac{\textbf{v}}{2}\right)=\frac{1}{2}\tilde \pi_1(\alpha+\textbf{v})$, which implies that, if we define the Klein disk of radius $1/2$ as
\begin{equation}
	\mathcal K_{1/2}=\{\textbf{x}/2\in\mathbb R^2,\ \Vert \textbf{x}\Vert^2< 1\},
\end{equation}
then the map
\begin{equation}
	\begin{array}{cccl}
		\varphi: & \mathcal K & \longrightarrow & \mathcal K_{1/2}  \\
		& (x_1,x_2)  & \longmapsto         & \varphi (x_1,x_2)=\frac{1}{2}(x_1,x_2),
	\end{array}
\end{equation}
is an isometry between $\mathcal K$ and $\mathcal K_{1/2}$, where the Riemannian metric on the latter is given by:
\begin{equation}
	ds^2_{\mathcal K_{1/2}}={(dx_1)^2+(dx_2)^2\over 1/4-(x_1^2+x_2^2)}+{(x_1dx_1+x_2dx_2)^2\over (1/4-(x_1^2+x_2^2))^2}\ .
\end{equation}

We recall that the geodesics of the Klein disk are extremely simple, being straight line segments, i.e. the chords inside the disk, contrary to those of the Poincaré disk, which are the diameters and the semicircles contained in the disk and orthogonal to its boundary. Moreover, the Klein distance on $\mathcal K_{1/2}$ coincides with the \textit{Hilbert distance}, defined as follows: let $p$ and $q$ be two interior points of the disk and let $r$ and $s$ be the two points of the boundary of the disk such that the segment $[r,s]$ contains the segment $[p,q]$. The Hilbert distance between $p$ and $q$ is defined by:
\begin{equation}
	d_H(p,q)={1\over 2}\log[r,p,q,s]\ ,
\end{equation}
where
\begin{equation}
	\label{eq:bir}
	[r,p,q,s]={\Vert q-r\Vert\over \Vert p-r\Vert}\cdot {\Vert p-s\Vert\over \Vert q-s\Vert}\ ,
\end{equation}
is the cross-ratio of the four points $r$, $p$, $q$ and $s$, for the proof see e.g. \cite{Beardon:99}. 
Without entering in details that would take too much space, we underline the importance of the Hilbert distance by saying that, in \cite{Berthier:21JMP}, this distance is shown to be intimately related to the relativistic Einstein-Poincaré addition law of the so-called perceptual chromatic vectors, which permits to link color perception to the formalism of special relativity in a rigorous way, bypassing the heuristic analysis of Yilmaz recalled in section \ref{sec:Yilmaz}.

\section{Discussion}\label{sec:discussion}

We have shown how Resnikoff's idea to abandon metameric classes of spectra and study the color space solely through its algebraic properties can be further refined by exploiting the properties of formally real Jordan algebras. This leads to a real quantum theory of color vision that permits to rigorously define colorimetric attributes and to understand chromatic opponency by means of quantum features. In this paper we have given a theoretical confirmation of the quantum nature of color perception by proving the existence of uncertainty relations satisfied by chromatic opponencies. 

The birth of quantum mechanics is related to the theoretical analysis of very simple experiments, such as the observation of interference pattern in polarized light or spin measurements by a pair of Stern–Gerlach devices. Analogously, this novel theory of color perception is developed from mathematical properties of vision gathered from experiments in extremely simple conditions. In the same way as quantum mechanics evolved into the much more complicated quantum-relativistic gauge field theory, we expect that a highly non-trivial extension of the model recalled in this paper is needed to understand color perception in more realistic conditions. For example, to deal with contextual effects as chromatic induction, one may try to replace the closed quantum system described here with open ones, as suggested in \cite{Berthier:2020}, or to extend the model to a field theory of color perception through bundles and connections, as suggested in \cite{Provenzi:16modernphysics,Provenzi:17}.

Another issue that we consider very intriguing is the deep comprehension of the concept of color measurement, which is intimately related to the definition of perceived color. As suggested in \cite{Berthier:2020}, the use of the so-called \textit{effects} in relation with generalized quantum observables and unsharp measurements may play a significant role in this analysis. Also the extension of our theory to encompass cognitive phenomena, as described in \cite{Gronchi:17}, seems to be a quite natural issue to explore.

On the geometrical side, it is quite remarkable that the set of quantum chromatic states can be embedded in the closure of the future lightcone $\overline{\mathcal L^+}$ via the projective tranformation (\ref{eq:Kleinhalf}), where it is identified with the Klein disk equipped with the Hilbert metric. This mathematical richness and clarity stands out even more if we compare it with the construction of the CIE $xy$ chromaticity diagram, which is also built through a projective transformations of the $XYZ$ CIE coordinates, namely $x=X/(X+Y+Z)$, $y=Y/(X+Y+Z)$ and $z=Z/(X+Y+Z)$. The flag-like shape of the CIE chromaticity diagram is far from having the regularity of a disk, moreover, it must be artificially closed with the purple line and, most importantly, it does not come naturally equipped with any metric. This last fact generated a lot of confusion when the well-known MacAdam ellipses \cite{MacAdam:42} were discovered and, eventually, when the `uniform' color spaces such as CIELab, CIELuv, etc. were build, the metric choice felt arbitrarily on the Euclidean distance. To cope with the non-Euclidean behavior of perceptual chromatic attributes, the coordinates of these spaces had to be artificially adjusted to fit the data with ad-hoc transformations and parameters. 

A firm aim of our future work is to avoid this kind of heuristic procedures by using only the minimal and more natural mathematical tools to extend the theory that we have exposed. As an example, it seems natural to investigate the quantum counterpart of the MacAdam ellipses through the analysis of the uncertainty relations that we have proven to be satisfied by the observables of our theory. The discovery of such features would be perfectly coherent with the quantum-like behavior of color perception.

\bibliographystyle{plain}
\bibliography{bibliography}

\end{document}